\documentclass{aastex}
\usepackage{spr-astr-addons}
\usepackage{url}
\usepackage{hyperref}
\hypersetup{colorlinks,citecolor=blue}
\RequirePackage{color}

\begin{document}
\title{Tolman IV fluid sphere in  bigravity}

\author{Ksh. Newton Singh \altaffilmark{}}
\altaffiltext{}{Department of Physics, National Defence Academy, Khadakwasla, Pune, Maharashtra-411023, India}
\altaffiltext{}{email: ntnphy@gmail.com}

\author{Susmita Sarkar \altaffilmark{}}
\altaffiltext{}{Department of Mathematics, Jadavpur University, Kolkata, India.}
\altaffiltext{}{email:susmita.mathju@gmail.com}

\author{Farook Rahaman \altaffilmark{}}
\altaffiltext{}{Department of Mathematics, Jadavpur University, Kolkata, India.}
\altaffiltext{}{email:rahaman@associates.iucaa.in}

\begin{abstract}
We present Tolman IV spacetime representing compact fluid sphere in bigravity. Here we have explored the effect of scale parameter $k$ in the local matter distribution of compact stars. We have model for three well-known compact stars and it shows that for lower values of $k$ leads to stiffer EoS. This claim is also supported by the graphical analysis. It can be observed that the sound speed and the adiabatic index are more for lower values of $k$. It is also seen that all the solutions of Einstein's field equations are still satisfying the field equations in the presence of a background metric $\gamma_{\mu \nu}$. However, the density and pressure does modified by extra term from the constant curvature background, thus affecting the EoS. One can also think that the parameter $\alpha \equiv 1/k^2$ as coupling constant between the $g_{\mu \nu}$ and $\gamma_{\mu \nu}$ and consequently more the coupling stiffer is the EoS. As $k\rightarrow \infty$, the background de-Sitter spacetime reduces to Minkowski's spacetime and the coupling vanishes. The solution satisfy the causality condition, all the energy conditions and  equilibrium under gravity and hydrostatic forces. The stability of the local stellar structure is enhanced by reducing the scalar curvature of the background spacetime. 

\end{abstract}

\section{Introduction}
A long years back, N. Rosen \citep{nr73,nr74,nr75,nr78} proposed a new modified theory to Einstein's general theory of relativity (GTR)  involving a background metric $\gamma_{\mu\nu}$ in addition to the usual physical metric $g_{\mu\nu}$, which is known as the ``{\it bimetric general theory of relativity} (BGTR  or bigravity)''. The background metric $\gamma_{\mu\nu}$  is governed by a parameter $k$.  In BGTR, the field equations are similar of Einstein's GTR however, with the ordinary derivatives of physical metrics are replaced by the covariant derivatives with respect to the background metric. Both $g_{\mu\nu}$ and $\gamma_{\mu\nu}$ are present in the field equations but only $g_{\mu\nu}$ interacts with matter. In the year 1978, \cite{nr78} initially written the static field equations in the form of Einstein's field equations with an additional bimetric term.  In the year 1985,  \cite{ah85} solved the field equations, which are obtained by using the procedure of \cite{nr80} corresponding to de-Sitter universe of constant curvature,  to obtain a model of compact star. They also proved that for an ordinary star, the results obtained from BGTR has no difference from the result obtained from Einstein's GTR. However, for a collapsed star model the two theories yield different results. The BGTR becomes a famous among the researcher within few years of the proposal and various aspects were investigated \citep{gs89,dr89,ig77,akrami}. It is investigated that the bimetric gravity has a subtle relationship with massive gravity \citep{vb12}. Studies of a finite size charge  and the static spherically symmetric field of an electric charge were done  in the framework of BGTR in \cite{fr79} and \cite{nr79}, respectively. In the year 1981, \cite{fd81} have investigated a charged point particle and then Rosen, himself discussed a classical model  for elementary particles in BGTR \citep{nr89}. \par

In the context of the Hassan-Rosen theory, spherically symmetric systems were first studied by \cite{comelli}, where in particular, the perturbative solutions to the equations of motion was published. In presence of bimetric gravity, \cite{Vol} performed an extensive numerical study, and gave conditions for the existence of asymptotically flat black hole solutions. Star solutions and the so-called Vainshtein mechanism was studied by \cite{Cri}. Solutions for charged black holes and for rotating black holes can be found in \cite{10}, \cite{11}, respectively. In massive bigravity, a general review of black holes was proposed by \cite{55}. \cite{edvard} proposed the phenomenology of stars and galaxies in massive bigravity by applying a parameter conditions for the existence of viable star solutions when the radius of the star is much smaller than the Compton wavelength of the graviton.\par

\cite{mr91} have studied on superdense celestial objects in generalized bimetric gravity with a variable gravitational constant.  In the framework of bimetric gravity, the solution for spherically symmetric self-gravitating anisotropic matter distribution was investigated by \cite{gs04} and  that solution agrees with the Einstein's GRT for a physical system compared to the solar system size of universe. A charged fluid was also  studied in BGTR \citep{sp10}. Recently,  \cite{ls18} have shown that the masses for superdense compact objects in bimetric theory can be essentially larger compared to the objects in GTR depending on the value of the bimetric parameter.\par

It is well known that, in the presence of an extra spin-2 field, bimetric theory describes gravitational interactions. Recently, Hassan et al. \citep{Has1,Has2} have proposed a particular bimetric theory (or bigravity) to avoid the ghost instability. This theory describes a nonlinear interactions of the gravitational metric with an additional spin-2 field. By assuming the Planck mass $M_f$ of the second metric, the ghost instability can be avoid by pushing back to early un-observably times. This limit the uses an effective cosmological constant in general relativity \citep{akrami}.\par

It is quite familiar that non-linear bimetric theories of gravity suffer from the same Boulware-Deser ghost instability \citep{deser}. To describe the interaction of gravity with a massive spin-2 meson, such theories were introduced. Recently, there has been renewed interest in bigravity due to their accelerating cosmological solutions \citep{kogan}. \cite{Has1} showed that the introduction of a kinetic term for the background metric in the ghost-free massive gravity leads to a bigravity theory which is also ghost free. A details study of cosmological solutions in bigravity can be found in refs. \cite{101,102,103}.\par

In present paper, we have developed a model of compact star in bimetric gravity where the metric potentials are chosen as Tolman IV metric potential. In one of our previous paper, we have investigated a new model of anisotropic compact star in (3+1)-dimensional spacetime. The model was obtained in the background of Tolman IV $g_{rr}$ metric potential as input and the field equatios were solved by assuming a suitable expression for radial pressure $p_r$ \citep{pnt}. The present paper is organized as follows: In sect.~2, field equations in bigravity is developed. In sect.~3 and 4, the model parameters are obtained by solving the field equations and physical properties are discussed, respectively. In sect.~5, we match our interior spacetime to the exterior Schwarzschild line element. Final two sections are devoted on a brief discussion.

\section{Field equations in bimetric gravity}

The bigravity modifies the Einstein's theory by accounting a background space-time. Therefore, one can defined the two symmetric tensor $g_{\mu \nu}$ and $\gamma_{\mu \nu}$ associated with the two spacetime and are defined as
\begin{equation}
ds_-^2 = g_{\mu \nu} dx^\mu dx^\nu~~~,~~~ds_b^2 = \gamma_{\mu \nu} dx^\mu dx^\nu.
\end{equation}
These two metric tensor are having non-vanishing determinants. Adopting the notation used by \cite{r12}, $\big\{\substack{\lambda \\ \mu~ \nu} \big\}$ is the Chistoffel symbol in $g-$different- iation and $\Gamma^\lambda_{\mu \nu}$ in $\gamma-$differentiation. The relationship between these two symbols are given as
\begin{equation}
\Big\{\substack{\lambda \\ \mu~~ \nu} \Big\} = \Gamma^\lambda_{\mu \nu}+\Delta^\lambda_{\mu \nu},
\end{equation} 
where, $\Delta^\lambda_{\mu \nu}$ is found to be related to $g_{\mu \nu}$ as
\begin{equation}
\Delta^\lambda_{\mu \nu} = {1 \over 2}~g^{\lambda \sigma} \Big( g_{\mu \sigma,\nu}+g_{\nu \sigma, \mu}-g_{\mu \nu ,  \sigma} \Big).
\end{equation}
Now the Riemann tensor related to the two metric functions can be written as
\begin{eqnarray}
\mathcal{R}^\lambda_{\mu \nu \sigma} &=& \mathcal{P}^\lambda_{\mu \nu \sigma} -\Delta^\lambda_{\mu \nu, \sigma}+\Delta^\lambda_{\mu \sigma, \nu}+\Delta^\lambda_{\alpha \nu} \Delta^\alpha_{\mu \sigma} -\Delta^\lambda_{\alpha \sigma} \Delta^\alpha_{\mu \nu} \nonumber \\
&=& -\Delta^\lambda_{\mu \nu, \sigma}+\Delta^\lambda_{\mu \sigma,\nu}+\Delta^\lambda_{\alpha \nu} \Delta^\alpha_{\mu \sigma} -\Delta^\lambda_{\alpha \sigma} \Delta^\alpha_{\mu \nu}. \label{eq4}
\end{eqnarray}
One can easily see that the tensor $\Delta^\lambda_{\mu \nu}$ is related to the normal Riemann curvature tensor. The curvature tensor $\mathcal{P}_{\lambda \mu \nu \sigma}$ of the constant curvature spacetime can be written as
\begin{equation}
\mathcal{P}_{\lambda \mu \nu \sigma}= {1 \over k^2} \Big( \gamma_{\mu \nu} \gamma_{\lambda \sigma} - \gamma_{\mu \sigma} \gamma_{\lambda \nu} \Big).
\end{equation}
By using (\ref{eq4}), one can also find the difference in the Ricci tensor i.e. $\mathcal{K}_{\mu \nu} = \mathcal{R}_{\mu \nu}-\mathcal{P}_{\mu \nu}$ which satisfy the same form of Einstein's field equations.

\begin{figure}[t]
\centering
\includegraphics[scale=.7]{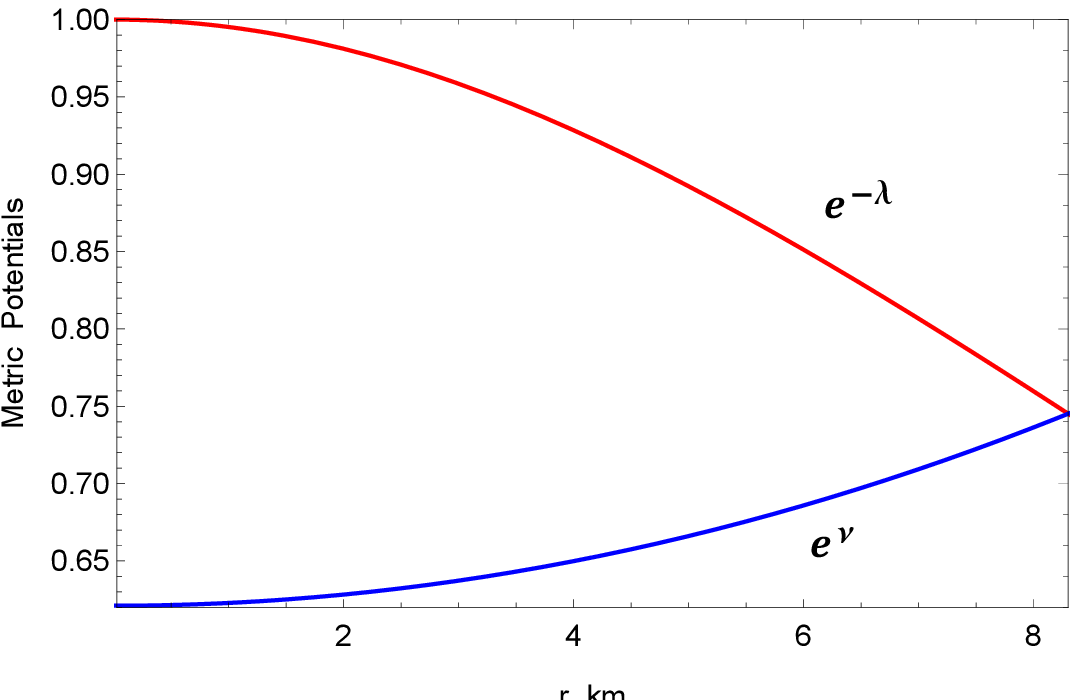}
\caption{Varaition of metric functions with respect to the radial coordinate $r$ for LMC X-4 ($M=1.04M_{\odot}$, $R=8.3km$ and $b = 0.001896,~ B = 0.788065)$ with $(a = 0.00266,~ k = 100)$, $(a = 0.0029,~ k = 150)$ and ($a = 0.003,~ k = 200$).}\label{f1}
\end{figure}

Finally, the field equations in bigravity can be written as \cite{nr78}
\begin{equation}
G_{\mu \nu} = S_{\mu \nu}-8\pi T_{\mu \nu},
\end{equation}where,
\begin{eqnarray}
G_{\mu \nu} &=& R_{\mu \nu}-{1 \over 2} R ~g_{\mu \nu} \\
S_{\mu \nu} &=& {3 \over k^2} \Big( \gamma_{\mu \nu}-{1 \over 2} g_{\mu \nu} g^{\alpha \beta} \gamma_{\alpha \beta} \Big)\\
T_{\mu \nu} &=& (\rho+p)u_\mu u_\nu -p~g_{\mu \nu}.
\end{eqnarray}

Now the interior and background de-Sitter space-time are taken as
\begin{eqnarray}
ds_-^2 &=& e^{\nu}dt^2-e^\lambda dr^2-r^2(d\theta^2+\sin^2 \theta ~ d\phi^2)\\
ds_b^2 &=& \left(1-{r^2 \over k^2}\right) dt^2-\left(1-{r^2 \over k^2} \right)^{-1}dr^2-r^2(d\theta^2 \nonumber \\
&& +\sin^2 \theta ~ d\phi^2).
\end{eqnarray}
Here $k$ represents the scale parameter in the de-Sitter universe.

Following \cite{r3} we can write the field equations as
\begin{eqnarray}
8\pi \rho &=& e^{-\lambda}\left( {\lambda' \over r}-{1 \over r^2}\right) +{1 \over r^2}+{3e^{-\nu} \over 2k^2} \label{rho1}\\
8\pi p &=& e^{-\lambda}\left( {\nu' \over r}+{1 \over r^2}\right) -{1 \over r^2}+{3e^{-\nu} \over 2k^2}\\
8\pi p &=& {e^{-\lambda} \over 2} \left[ \nu''+{\nu'^2 \over 2}+{\nu'-\lambda' \over r}-{\nu' \lambda' \over r}\right]+{3e^{-\nu} \over 2k^2}. \label{p1}
\end{eqnarray}
Due to the constant curvature of the background metric, the density and pressure are modified as 
\begin{eqnarray}
\rho_e (r) &=& \rho(r)-{3e^{-\nu} \over 16\pi k^2} \\
p_e(r) &=& p(r) -{3e^{-\nu} \over 16\pi k^2}
\end{eqnarray}
which also satisfies the TOV-equation i.e.
\begin{equation}
-{\nu' \over 2}(\rho_e+p_e)-{dp_e \over dr}=0.
\end{equation}

To analyze in deeper aspects, we will ansatz the Tolman IV spacetime and discuss its behavior w.r.t. the scale factor $k$.

\begin{figure}[t]
\centering
\includegraphics[scale=.7]{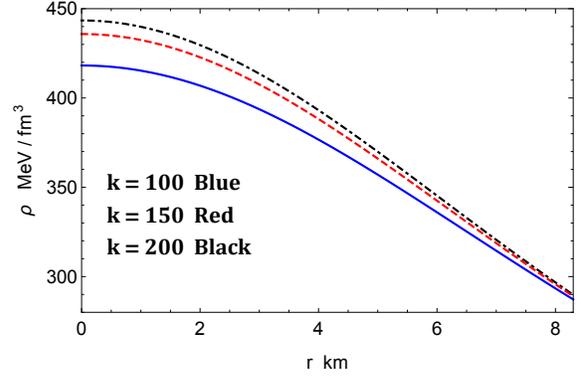}
\caption{Variation of density with respect to the radial coordinate $r$ for LMC X-4 ($M=1.04M_{\odot}$, $R=8.3km$ and $b = 0.001896,~ B = 0.788065)$ with $(a = 0.00266,~ k = 100)$, $(a = 0.0029,~ k = 150)$ and ($a = 0.003,~ k = 200$).}\label{f2}
\end{figure}

\section{Tolman IV solution in bigravity}

The Tolman IV \citep{tol1} spacetime is given by
\begin{eqnarray}
ds^2_- &=& B^2 (1+ar^2) dt^2-\frac{2 a r^2+1}{\left(a r^2+1\right) \left(1-b r^2\right)}~dr^2 \nonumber \\
&& -r^2(d\theta^2+\sin ^2 \theta ~d\phi^2).
\end{eqnarray}
The nature of the metric functions are shown in Fig. \ref{f1}.

\begin{figure}[t]
\centering
\includegraphics[scale=.7]{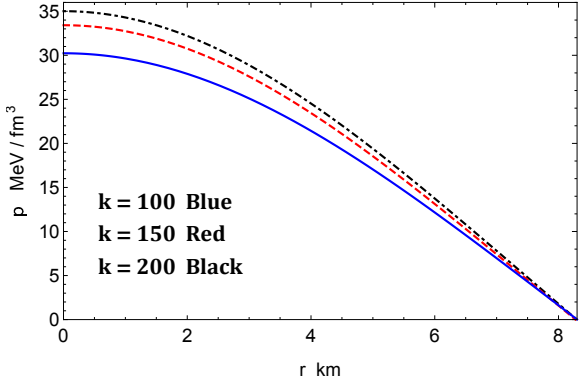}
\caption{Variation of pressure with respect to the radial coordinate $r$ for for LMC X-4 ($M=1.04M_{\odot}$, $R=8.3km$ and $b = 0.001896,~ B = 0.788065)$ with $(a = 0.00266,~ k = 100)$, $(a = 0.0029,~ k = 150)$ and ($a = 0.003,~ k = 200$).}\label{f3}
\end{figure}

For the Tolman IV space-time the density and pressure of the stellar configuration can be written as
\begin{eqnarray}
\rho (r) &=& \frac{1}{16 B^2 k^2 \left(2 a r^2+1\right)^2 \left(\pi  a r^2+\pi \right)} \bigg[ 4 a^3 B^2 k^2 r^4 \nonumber \\
&&\left(3 b r^2+1\right)+2 a^2 \big\{B^2 k^2 r^2 \left(13 b r^2+5\right) +6 r^4\big\} \nonumber \\
&& +2 a \left\{B^2 k^2 \left(10 b r^2+3\right)+6 r^2\right\}+6 b B^2 k^2+3 \bigg] \nonumber \\ 
\end{eqnarray}
\begin{eqnarray}
p(r) &=& \frac{1}{8 \pi } \bigg[\frac{3}{2 B^2 k^2 \left(a r^2+1\right)}-\frac{\left(3 a r^2+1\right) \left(b r^2-1\right)}{2 a r^4+r^2} \nonumber \\
&& -\frac{1}{r^2} \bigg].
\end{eqnarray}

The graphical representations of our obtained density and pressure are shown in Figs. \ref{f2} and \ref{f3} with respect to the radial coordinate $r$ for the compact star LMC X-4. Fig. 

The density and pressure gradients are calculated as
\begin{eqnarray}
{d\rho \over dr} &=& - \frac{ar}{8 \pi  B^2 k^2 \left(a r^2+1\right)^2 \left(2 a r^2+1\right)^3} \bigg[8 a^4 B^2 k^2 r^6 \nonumber \\
&& +4 a^3 \big\{B^2 k^2 r^4 (b r^2+9) +6 r^6\big\}+ 6 a^2 \big\{B^2 k^2 r^2 \nonumber \\
&& (3 b r^2+8)+6 r^4\big\}+2 a \big\{2 B^2 k^2 (6 b r^2+5)+ \nonumber \\
&& 9 r^2\big\} +10 b B^2 k^2+3\bigg]\\
{dp \over dr} &=& -\frac{a r}{8 \pi  B^2 k^2 \left(a r^2+1\right)^2 \left(2 a r^2+1\right)^2} \bigg[4 a^3 B^2 k^2 r^4 \nonumber \\
&& +2 a^2 \big\{B^2 k^2 r^2 (b r^2+4) +6 r^4\big\}+4 a \big\{B^2 k^2 \nonumber \\
&& (b r^2+1)+3 r^2\big\}+2 b B^2 k^2+3 \bigg].
\end{eqnarray}

\begin{figure}[t]
\centering
\includegraphics[scale=.7]{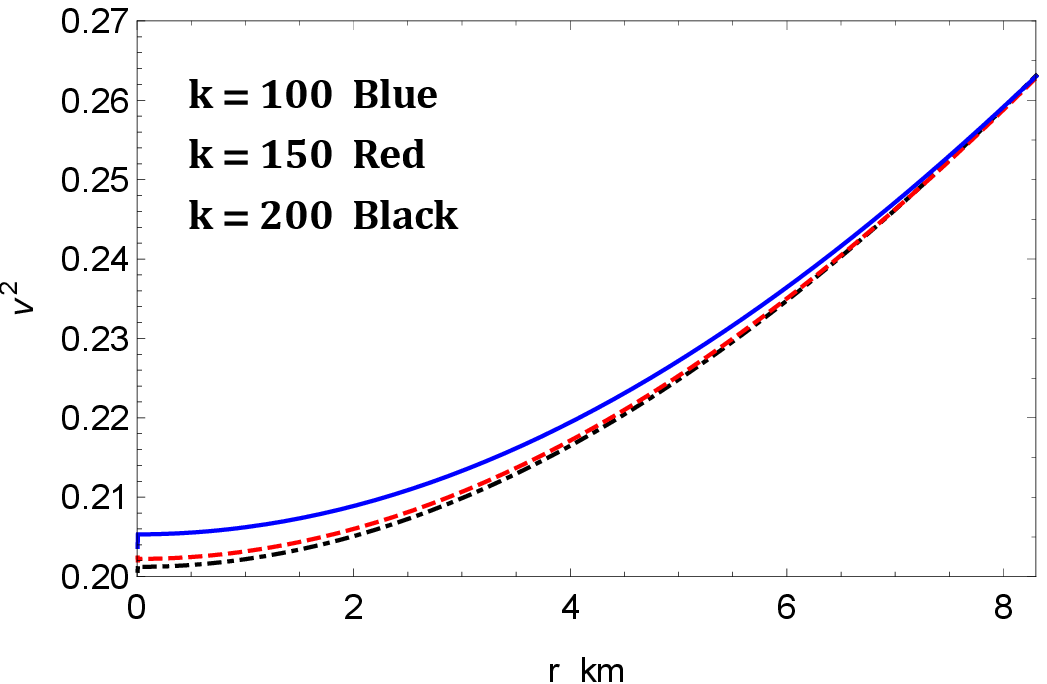}
\caption{Varaition of sound speed with respect to the radial coordinate $r$ for LMC X-4 ($M=1.04M_{\odot}$, $R=8.3km$ and $b = 0.001896,~ B = 0.788065)$ with $(a = 0.00266,~ k = 100)$, $(a = 0.0029,~ k = 150)$ and ($a = 0.003,~ k = 200$).}\label{f4}
\end{figure}

The causality condition can be verify by computing the speed of sound as
\begin{eqnarray}
v^2 &=& \frac{dp}{d\rho} = (2 a r^2+1) \Big[4 a^3 B^2 k^2 r^4+2 a^2 \big\{B^2 k^2 r^2 \nonumber \\
&& (b r^2+4)+6 r^4\big\}+4 a \big\{B^2 k^2 (b r^2+1) +3 r^2\big\} + \nonumber \\
&& 2 b B^2 k^2+3\Big] \Big[8 a^4 B^2 k^2 r^6+4 a^3 \big\{B^2 k^2 r^4  (b r^2+9) \nonumber \\
&& +6 r^6\big\}+6 a^2 \big\{B^2 k^2 r^2(3 b r^2+8)+6 r^4\big\}+2 a   \nonumber \\ 
&& \big\{2 B^2 k^2(6 b r^2+5)+9 r^2\big\}+10 b B^2 k^2+3 \Big]^{-1}.
\end{eqnarray}
To satisfy the causality condition the speed of sound $v$ should be less than unity (in $c=1=G$ unit). The trend of the sound speed is shown in Fig. \ref{f4}.

To analyze the energy conditions we can calculate the following parameters:
\begin{eqnarray}
\rho(r)-p(r) &=& \frac{6 a^2 b r^4+6 a b r^2+a+2 b}{4 \pi (2 a r^2+1)^2}\label{ec1}
\\
\rho(r)-3p(r) &=& \frac{1}{8 B^2 k^2 (2 a r^2+1)^2 (\pi  a r^2+\pi)} \bigg[2 b B^2  \nonumber \\
&& k^2(12 a^3 r^6+23 a^2 r^4+ 14 a r^2+3) \nonumber \\
&& -4 a^3B^2 k^2 r^4-4 a^2(B^2 k^2 r^2+3 r^4)\nonumber \\
&& -12 a r^2-3 \bigg]\label{ec2}.
\end{eqnarray}

For any physical matters, the strong, weak, dominant and null energy conditions need to be satisfy i.e.
\[\rho \ge 0;~~\rho-p \ge 0;~~\rho-3p \ge 0;~~ \rho \ge |p|.\]

\begin{figure}[t]
\centering
\includegraphics[scale=.7]{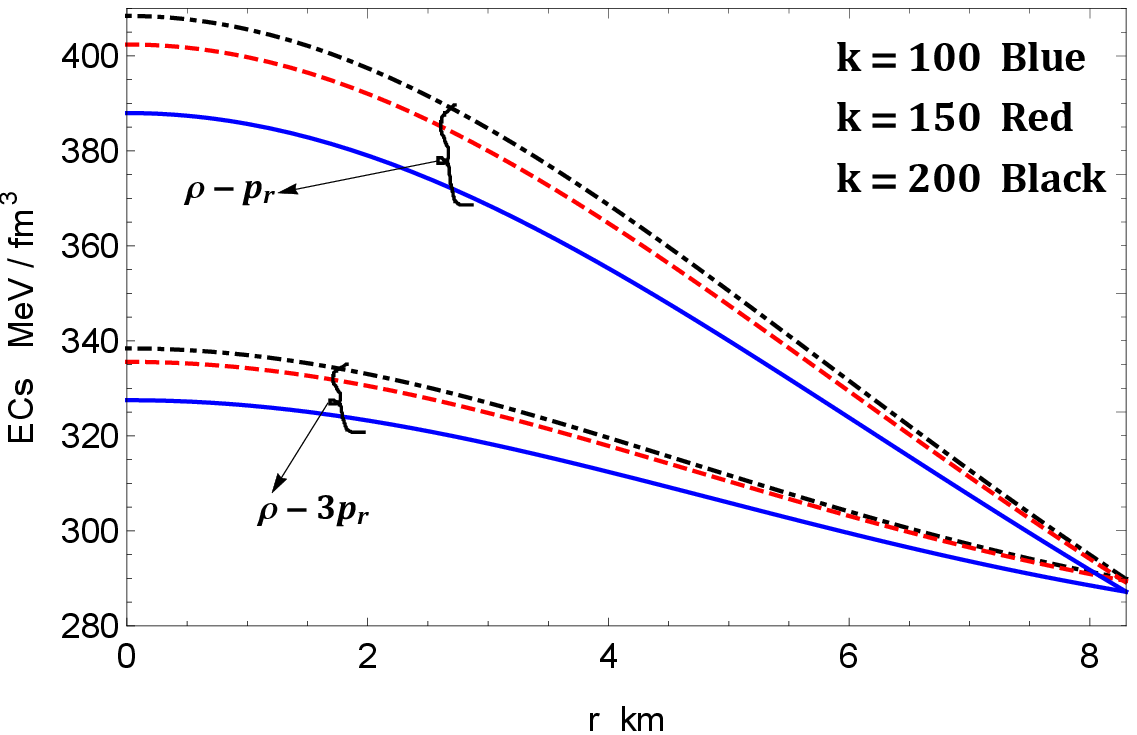}
\caption{Varaition of energy conditions with respect to the radial coordinate $r$ for LMC X-4 ($M=1.04M_{\odot}$, $R=8.3km$ and $b = 0.001896,~ B = 0.788065)$ with $(a = 0.00266,~ k = 100)$, $(a = 0.0029,~ k = 150)$ and ($a = 0.003,~ k = 200$).}\label{f5}
\end{figure}

The results obtained in Eqs. (\ref{ec1}) and (\ref{ec2}) are shown graphically in Fig. \ref{f5}.

\begin{figure}[t]
\centering
\includegraphics[scale=.7]{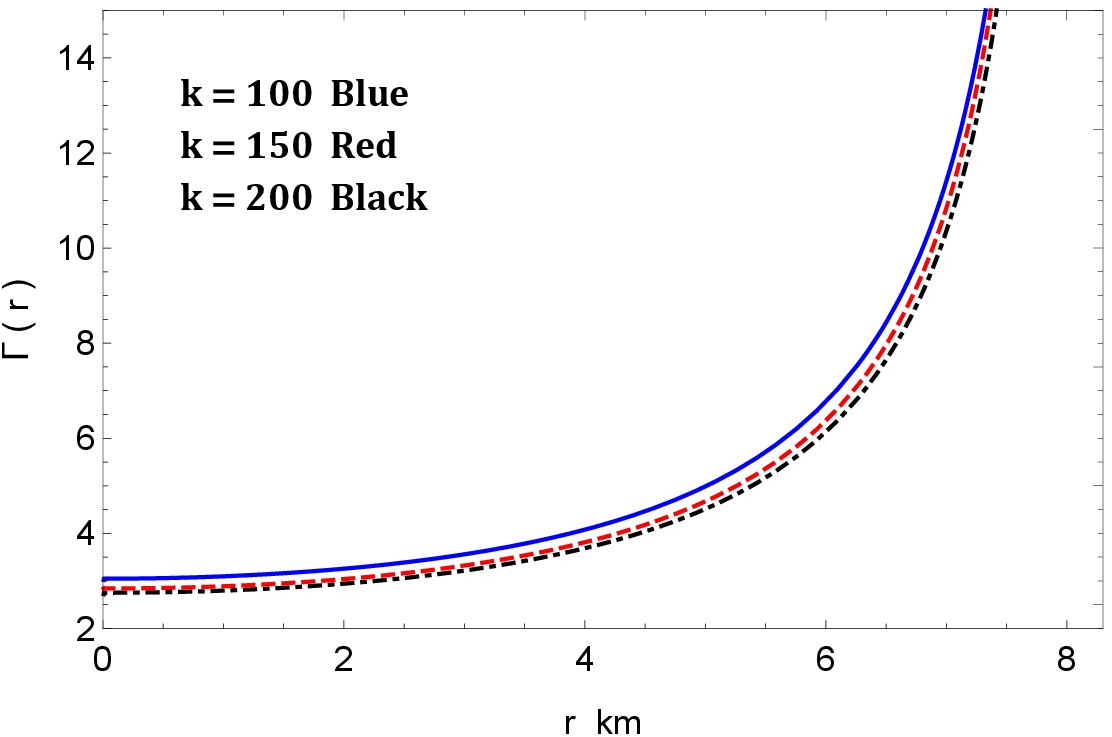}
\caption{Varaition of adiabatic index with respect to the radial coordinate $r$ for LMC X-4 ($M=1.04M_{\odot}$, $R=8.3km$ and $b = 0.001896,~ B = 0.788065)$ with $(a = 0.00266,~ k = 100)$, $(a = 0.0029,~ k = 150)$ and ($a = 0.003,~ k = 200$).}\label{f6}
\end{figure}

\begin{figure}[t]
\centering
\includegraphics[scale=.7]{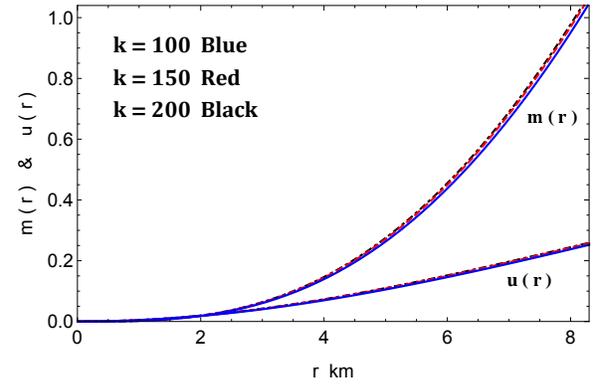}
\caption{Varaition of mass function and compactness parameter with respect to the radial coordinate $r$ for LMC X-4 ($M=1.04M_{\odot}$, $R=8.3km$ and $b = 0.001896,~ B = 0.788065)$ with $(a = 0.00266,~ k = 100)$, $(a = 0.0029,~ k = 150)$ and ($a = 0.003,~ k = 200$).}\label{f7}
\end{figure}

Now we can defined the adiabatic index, mass function and compactness parameter as
\begin{eqnarray}
\Gamma &=& {\rho+p \over p} ~{dp \over d\rho}\\
m(r) &=& 4\pi \int_0^r \rho(r) r^2 dr = \frac{1}{4 a^{3/2} B^2 k^2 (2 a r^2+1)} \nonumber \\
&& \bigg[ \sqrt{a} r \big\{2 a^2 B^2 k^2 r^2 (b r^2+1)+ 2 a r^2 (b B^2 k^2+3) \nonumber \\
&& +3\big\}-3(2 a r^2+1) \tan ^{-1}(\sqrt{a} r)\bigg] 
\end{eqnarray}
\begin{eqnarray}
u(r) &=& {2m(r) \over r} =  \frac{1}{2 a^{3/2} B^2 k^2(2 a r^3+r)} \nonumber \\
&& \bigg[\sqrt{a} r \big\{2 a^2 B^2 k^2 r^2 (b r^2+1)+2 a r^2(b B^2 k^2 \nonumber \\
&& +3)+3\big\}-3(2 a r^2+1) \tan ^{-1}(\sqrt{a} r) \bigg].
\end{eqnarray}
The variation of the above physical quantities can be seen in Figs. \ref{f6} and \ref{f7}.

\section{Physical properties of the solution}
The non-singular nature of the solution can be seen from the central values of density and pressure. The central values of these parameter can be written as
\begin{eqnarray}
\rho_c &=&  \frac{6 a B^2 k^2+6 b B^2 k^2+3}{16 \pi  B^2 k^2}  > 0 \label{c1}\\
p_c &=& \frac{2 a-2 b+3/B^2 k^2}{16 \pi } >0. \label{c2}
\end{eqnarray}

For any physical matter, the Zeldovich's condition \citep{zel} has to be satisfy i.e. $p_c / \rho_c \le 1$, which implies
\begin{eqnarray}
0 \le a+2b.
\end{eqnarray}

The TOV-equation of hydrostatic equilibrium can be written as
\begin{eqnarray}
 -{\nu' \over 2} (\rho+p)-{dp \over dr} &=& 0 \nonumber \\
\mbox{or} ~~~F_g+F_h&=&0.
\end{eqnarray}

\begin{figure}[t]
\centering
\includegraphics[scale=.7]{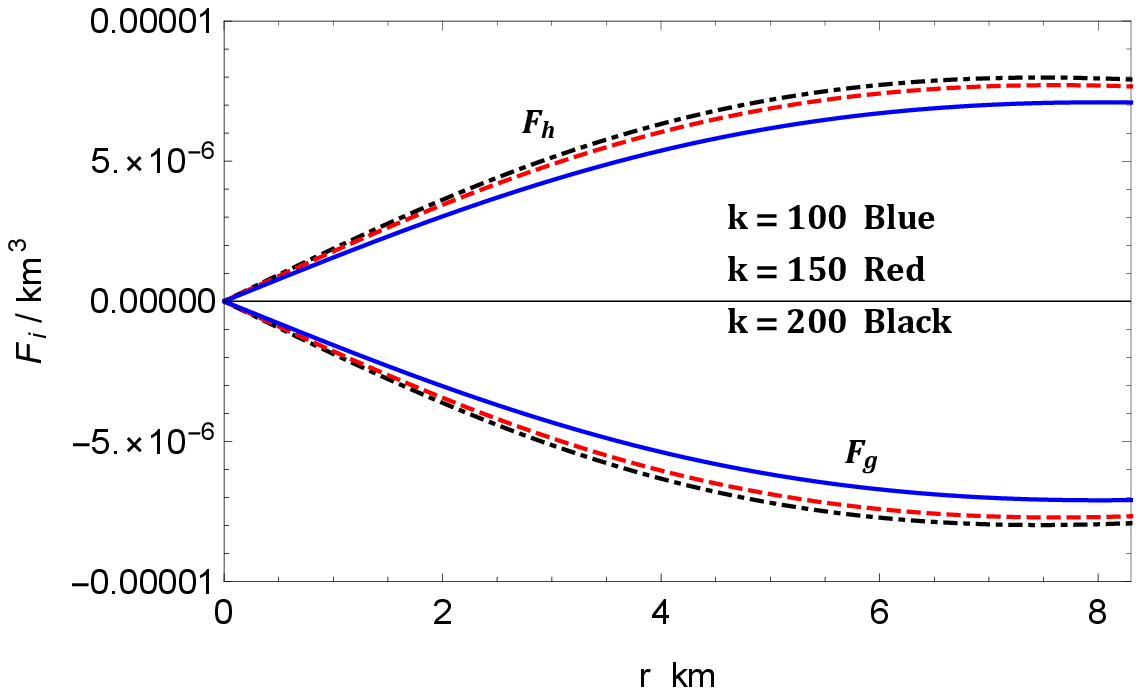}
\caption{Varaition of various forces in TOV-equation with respect to the radial coordinate $r$ for LMC X-4 ($M=1.04M_{\odot}$, $R=8.3km$ and $b = 0.001896,~ B = 0.788065)$ with $(a = 0.00266,~ k = 100)$, $(a = 0.0029,~ k = 150)$ and ($a = 0.003,~ k = 200$).}\label{f8}
\end{figure}

Here the gravitational and hydrostatic forces can be defined as
\begin{eqnarray}
F_g &=& -{\nu' \over 2} (\rho + p)= \frac{-a r}{8 \pi  B^2 k^2 \left(a r^2+1\right)^2 \left(2 a r^2+1\right)^2} \nonumber \\
&& \bigg[3+4 a^3 B^2 k^2 r^4 +2 a^2 \big\{B^2 k^2 r^2 (b r^2+4)  +6 r^4\big\} \nonumber \\
&& +4 a \big\{B^2 k^2 (b r^2+1) +3 r^2\big\}+2 b B^2 k^2 \bigg] \\
F_h &=& -{dp \over dr}= \frac{a r}{8 \pi  B^2 k^2 \left(a r^2+1\right)^2 \left(2 a r^2+1\right)^2} \bigg[4 a^3 B^2 \nonumber \\
&& k^2 r^4+2 a^2 \big\{B^2 k^2 r^2  \nonumber \\
&& (b r^2+4)+6 r^4\big\}+4 a \left\{B^2 k^2 (b r^2+1)+3 r^2\right\} \nonumber \\
&& +2 b B^2 k^2+3 \bigg].
\end{eqnarray}
The exact profiles of the gravitational force $F_g$ and hydrostatics force $F_h$ for our model are displayed in Fig. \ref{f8}, which indicates that solution represents the equilibrium matter configuration.

\begin{figure}[t]
\centering
\includegraphics[scale=.7]{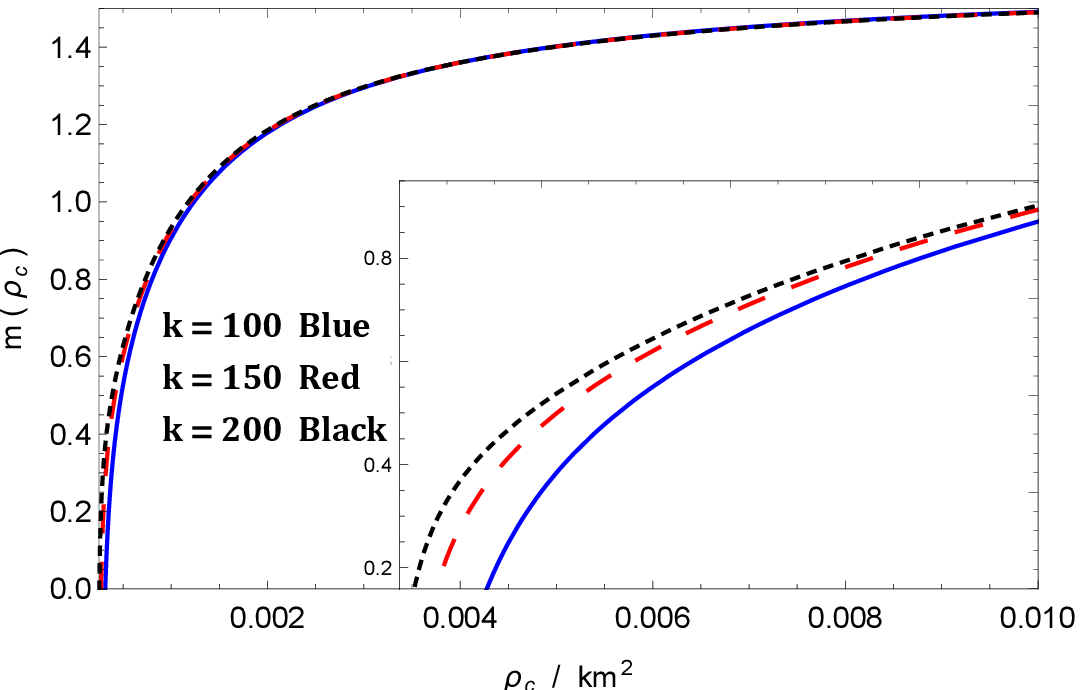}
\caption{Varaition of mass with respect to the radial central density $\rho_c$ for LMC X-4 ($M=1.04M_{\odot}$, $R=8.3km$ and $b = 0.001896,~ B = 0.788065)$ with $(a = 0.00266,~ k = 100)$, $(a = 0.0029,~ k = 150)$ and ($a = 0.003,~ k = 200$).}\label{f9}
\end{figure}

To check the stability of the solution and how the parameter $k$ affect it, one can adopt the static stability criterion. As per this criterion, the mass of the system must be increasing function of its central density i.e. $\partial m / \partial \rho_c > 0$ or otherwise unstable. This insures the stability of the system under radial perturbations. The mass as a function of its central can be given as
\begin{eqnarray}
m(\rho_c) &=& {R \over 4} \Bigg[1+b r^2-\frac{18}{2 B^2 k^2 (3 b-8 \pi  \rho )+3}- \nonumber \\
&& \frac{3\left(b r^2-1\right)}{\left[2 r^2 (3 b-8 \pi  \rho_c )-3\right]+3 r^2 / B^2 k^2}\Bigg] \nonumber \\
&& -\frac{9 B^2 k^2 \sqrt{24 \pi  \rho_c -9 b-9/2 B^2 k^2}}{\left[2 B^2 k^2 (3 b-8 \pi  \rho_c )+3\right]^2}  \nonumber \\
&& \tan ^{-1}\left(r \sqrt{\frac{8 \pi  \rho_c }{3}-b-\frac{1}{2 B^2 k^2}}\right).
\end{eqnarray} 
The variation of mass with central density is shown in Fig. \ref{f9}. It signifies that the solution gain its stability when $k$ increases i.e. whenever the scalar curvature of the background de-Sitter spacetime ($R=12/k^2$) deceases, the stability of the local stellar system is enhanced. This is because, as $k$ increases the stable range of density increases thereby the change in density during radial oscillations does not trigger gravitational collapse.

\section{Matching of interior and exterior boundary}

Assume the exterior metric as the Schwarzschild vacuum for the region $r>R$ i.e.
\begin{eqnarray}
ds_+^2 &=& \left(1-{2M \over r}\right) dt^2-\left(1-{2M \over r}\right)^{-1}dr^2 \nonumber \\
&& -r^2(d\theta^2+\sin^2 \theta ~d\phi^2).
\end{eqnarray}

Matching at the boundary $r=R$ we get
\begin{eqnarray}
e^{\nu(R)} &=& e^{-\lambda(R)} = 1-{2M \over R} \label{bo1}
\\
p(R) &=& 0 \label{bo2}.
\end{eqnarray}

\begin{figure}[t]
\centering
\includegraphics[scale=.7]{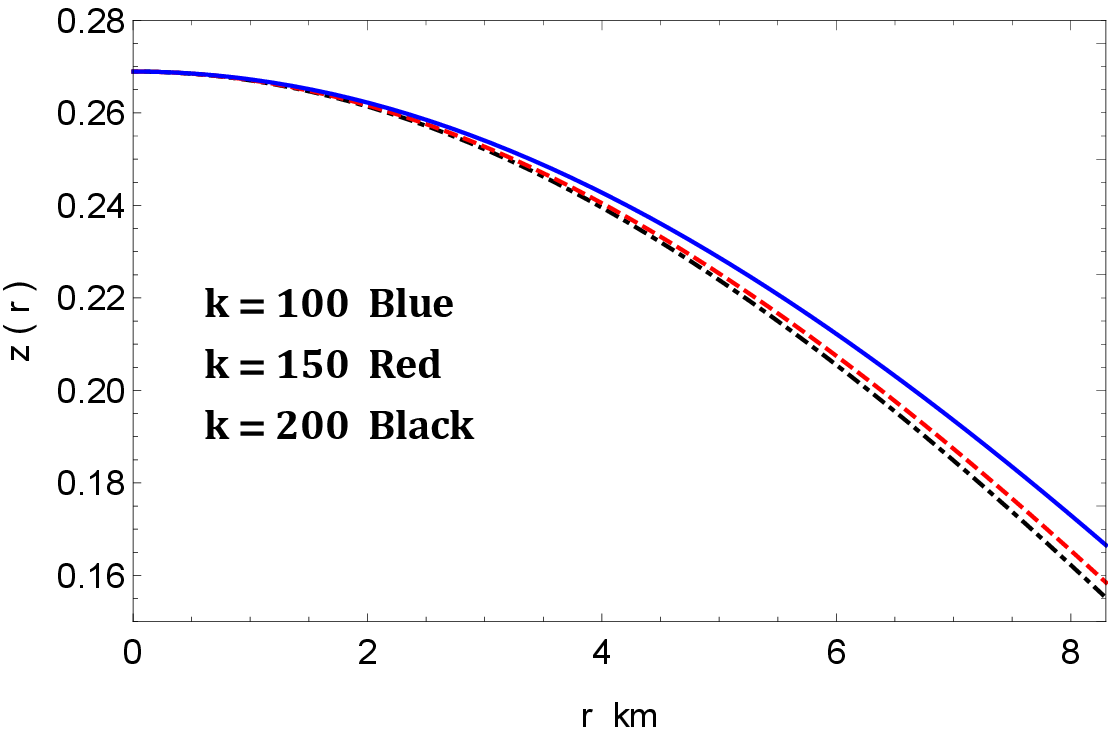}
\caption{Varaition of redshift with respect to the radial coordinate $r$ for LMC X-4 ($M=1.04M_{\odot}$, $R=8.3km$ and $b = 0.001896,~ B = 0.788065)$ with $(a = 0.00266,~ k = 100)$, $(a = 0.0029,~ k = 150)$ and ($a = 0.003,~ k = 200$).}\label{f10}
\end{figure}

\begin{figure}[t]
\centering
\includegraphics[scale=.7]{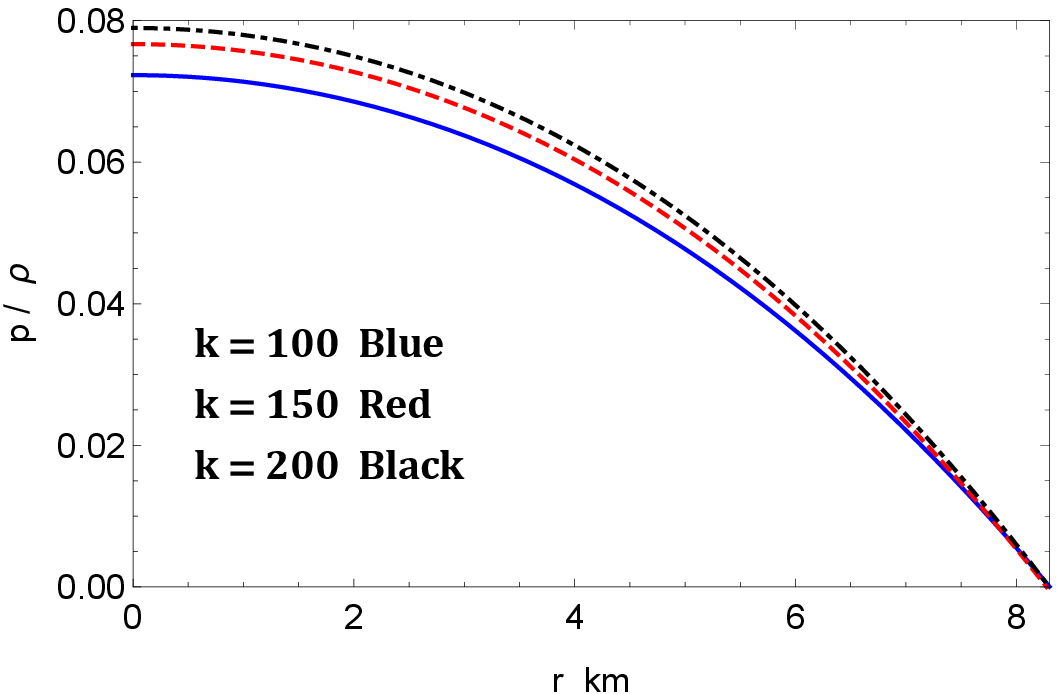}
\caption{Varaition of equation of state parameter with respect to the radial coordinate $r$ for LMC X-4 ($M=1.04M_{\odot}$, $R=8.3km$ and $b = 0.001896,~ B = 0.788065)$ with $(a = 0.00266,~ k = 100)$, $(a = 0.0029,~ k = 150)$ and ($a = 0.003,~ k = 200$).}\label{f11}
\end{figure}

On using the boundary conditions (\ref{bo1}) and (\ref{bo2}) we get
\begin{eqnarray}
a &=& \frac{b R^3-2 M}{R^2 (4 M-b R^3-R)} \\
b &=& \frac{2 a^2 B^2 k^2 R^2+2 a B^2 k^2+6 a R^2+3}{2 B^2 k^2 (a R^2+1) (3 a R^2+1)} \\
B &=& \sqrt{\frac{R-2 M}{R \left(a R^2+1\right)}}
\end{eqnarray}
Here we have used $M,~R$ and $k$ as free parameters and rest of the constants are found using the above three equations.

Now the red-shift and equation of state parameter of compact stars can be determined as
\begin{eqnarray}
z(r) &=& e^{-\nu/2}-1 = {1 \over B \sqrt{1+ar^2}}-1  \\
\omega &=& {p \over \rho} \le 1.
\end{eqnarray}
Fig. \ref{f10} and \ref{f11} clarifies the behavior of redshift function and the equation of state parameter with respect to the radial coordinate function $r$ respectively.

\begin{table*}[!t]
\caption{All the parameters corresponds to well-behaved solution representing few well-known compact stars.}\label{tab2}
\centering
\begin{tabular}{|c|c|c|c|c|c|c|c|c|}
\hline
Objects & $a ~km^{-2}$ & $b~km^{-2}$ & $B$ & $k$ & $M/M_\odot$ & $R~km$ & $M_{obs}/M_\odot$ & $R_{obs}~km$ \\
\hline
SMC X-4 & 0.003 & 0.0020484 &  0.757390 & 80 & 1.29 & 8.831 & 1.29$\pm$0.05 & 8.831$\pm$0.09   \\
LMC X-4 & 0.003 & 0.0018955 &  0.788065 & 200 & 1.04 & 8.3 & 1.04$\pm$0.09 & 8.301$\pm$0.2   \\
PSR J1614-2230 & 0.002 & 0.00332169 &  0.706809 & 33 & 1.97 & 9.69 & 1.97$\pm$0.04 & 9.69$\pm$0.2   \\
\hline
\end{tabular}
\end{table*}

\begin{table*}[!t]
\caption{All the parameters corresponds to well-behaved solution representing few well-known compact stars.}\label{tab1}
\centering
\begin{tabular}{|c|c|c|c|c|c|c|c|c|}
\hline
Objects & $u$ & $z_c$ & $z_s$ & $\rho_c$ & $\rho_s$ & $p_c$ & $\Gamma_c$  \\
& & & & $g/cm^3$ & $g/cm^3$ & $dyne/cm^2$ &  \\
\hline
SMC X-4 & 0.292 & 0.32 & 0.19 & 8.34 $\times 10^{14}$ & 5.30$\times 10^{14}$ & 6.54$\times 10^{34}$ & 2.592    \\
LMC X-4 & 0.251 & 0.269 & 0.155 & 7.89$\times 10^{14}$ & 5.16$\times 10^{14}$ & 5.59$\times 10^{34}$ & 2.726   \\
PSR J1614-2230 & 0.407 & 0.415 & 0.298 & 10.03$\times 10^{14}$ & 7.40$\times 10^{14}$ & 6.99$\times 10^{34}$ & 3.601    \\
\hline
\end{tabular}
\end{table*}

\section{Results and discussions}

In this article, we have explored the behavior of the Tolman IV solution in bimetric gravity describing relativistic fluid sphere. To demonstrate the effect of the scale parameter $k$ on the solution we used graphical methods for a specific compact star i.e. LMC X-4. In the framework of bigravity as well, the Tolman IV spacetime behaves in good agreement as in GR. 

The obtained physical parameters, density and pressure are as expected positive, maximum at the centre and then monotonically decreasing in nature towards the surface , Figs. \ref{f1} (Right) and \ref{f2} (Left). The mass and the compactness parameter are monotonically increasing in nature toward the surface, shown in Fig. \ref{f4} (Left). Therefore, all the physical parameters are well-behaved and physically acceptable. satisfies the TOV-equation even in the presence of the background spacetime, (see Fig. \ref{f4}, Right). Hence, the solution can represent static and equilibrium relativistic fluid spheres. The nature of the equation of state parameter (EoS) i.e. $p/\rho <1$ indicates the solution represents the physically acceptable matter distribution \citep{fr10}. 

The scale parameter $k$ can affect the solution and thus affect the corresponding equation of state (EoS) or the internal compositions of the compact fluid object. We have provided the central values of some physical quantities in Table \ref{tab1}, whereas the Table-\ref{tab2} display the compatible constant parameter to model SMC X-4, LMC X-1 and PSR J1614-2230. As we can see, smaller value of $k$ corresponds to stiffer EoS i.e. $k = 33$ along with the given parameters, corresponds to mass and radius of PSR J1614-2230 ($1.97M_{\odot},~9.69km$) with compactness parameter of 0.407 and for $k=200$ corresponds to LMC X-4 ($1.04M_\odot,~8.3km$) with $u=0.251$. This claim is further supported by the graphical representations. The velocity of sound and central values adiabatic index are  more for $k=100$ than $k=200$ (see Figs. \ref{f2} Right and \ref{f3} Right). This suggest that as $k$ increases the stiffness of the solution decreases.

Furhter, the solution also satisfy the causality and all the energy conditions (see Figs. \ref{f2} Right, \ref{f3} Left). Consequently, the solution representing matter distribution is physical. The stability of compact star is one of the most vital requirement, for that, we have focused to discuss the stability with respect to the variation of  adiabatic index $\Gamma$   inside the compact star. For a Newtonian fluid, stable configuration can be achieved if $\Gamma \ge 4/3$ \citep{hb64}. In our model, the stable nature of the configuration can be convinced from the graphical representation of the $\Gamma$, provided in Fig. \ref{f3} (Right). Further, the solution also satisfies the static stability criterion showing that the stability is enhance when the curvature of the background metric decreases (Fig. \ref{f5}). Therefore, the stability will be maximum when assumed a flat or Minkowski's spacetime.

\section{Conclusion}

Since the local spacetime interior to the compact star is part of the global structure in the universal scale, the physics of such object can be influence by the scale parameter of the universe. Even though the effect of the scale parameter may be small, however, the physics of such small perturbations can be interesting and worth complete analysis. \cite{ah85} have shown that there is possible to exist a configuration in hydrostatic equilibrium  for a collapsing star filling its Schwarzschild sphere when no such configuration exist in Einstein's gravity. It is also mentioned that in bigravity even leads a cosmology without a ``Big Bang" \citep{nr80}.

\cite{nr80} also solved the field equations with background metric near the Schwarzschild sphere and found that the field differs from that of the Einstein's GTR. Instead of a black hole they have obtained an impenetrable sphere. It is also found that the bigravity is identical with the ordinary general relativity at large distances (scale of solar system), however, at very large distances (scale of clusters of galaxies) the scale parameter affect the field equations and thus the dynamics.

As mentioned by \cite{ah85}, all the solutions of field equations in general relativity does satisfies the field equations with a background and the same is also supported by the current work with more details analysis. The internal structure of compact stars can also be influence by the background metric. We have shown that the local structures can also be influence by the background spacetime.  Hence, the bimetric GTR can inspire many researchers and also can contribute to new physics in future.

\section*{Acknowledgement}
Farook Rahaman would like to thank the authorities of the Inter-University Centre for Astronomy and Astrophysics, Pune, India for providing research facilities.  Susmita Sarkar is grateful to UGC (Grant No.: 1162/(sc)(CSIR-UGC NET , DEC 2016)), Govt. of India, for financial support.

\section*{Conflict of interest}
The authors declare no conflict of interest.

\end{document}